\documentclass[iop]{emulateapj}

\usepackage[]{algorithm}
\usepackage{algpseudocode}
\usepackage{amssymb}

\usepackage{graphicx}
\usepackage{amsmath}
\usepackage{longtable}
\usepackage{color}

\definecolor{Red}{rgb}{1,0,0}
\definecolor{Green}{rgb}{0,1,0}
\definecolor{forestgreen}{rgb}{0.13, 0.55, 0.13}
\definecolor{Blue}{rgb}{0,0,1}

\newcommand{\ie}{i.e.,\,}
\newcommand{\be}{\begin{equation}}
\newcommand{\ee}{\end{equation}}
\newcommand{\bea}{\begin{eqnarray}}
\newcommand{\eea}{\end{eqnarray}}

\newcommand{\etal}{et al.}

\newcommand{\nhat}{\hat{\bf n}}

\begin{document}

\bibliographystyle{apsrev}

\title{ALGORITHMS AND PROGRAMS FOR STRONG GRAVITATIONAL LENSING IN KERR SPACE-TIME INCLUDING POLARIZATION
}

\author{Bin Chen\altaffilmark{1,2}, Ronald Kantowski\altaffilmark{2}, Xinyu Dai\altaffilmark{2}, Eddie Baron\altaffilmark{2}, Prasad Maddumage\altaffilmark{1}}

\altaffiltext{1}{Research Computing Center, Department of Scientific Computing,
Florida State University, Tallahassee, FL 32306, USA, bchen3@fsu.edu}

\altaffiltext{2}{Homer L. Dodge Department of Physics and Astronomy,
University of Oklahoma, Norman, OK 73019, USA}

\begin{abstract}
Active galactic nuclei (AGNs) and quasars are important astrophysical objects to understand.
Recently, microlensing observations have constrained the size of the quasar X-ray emission region to be of the order of 10 gravitational radii of the central supermassive black hole.
For distances within a few gravitational radii, light paths are strongly bent by the strong gravity field of the central black hole.
If the central black hole has nonzero angular momentum (spin), a photon's polarization plane will be rotated by the gravitational Faraday effect.
The observed X-ray flux and polarization will then be influenced significantly by the strong gravity field near the source.
Consequently, linear gravitational lensing theory is inadequate for such extreme circumstances.
We present  simple algorithms computing strong lensing effects of Kerr black holes, including effects on polarization.
Our algorithms are realized in a program ``KERTAP" in two versions: MATLAB and Python. 
The key ingredients of KERTAP are:  a graphic user interface,  a {\it backward} ray-tracing algorithm,  a polarization propagator dealing with gravitational Faraday rotation, and algorithms computing observables such as flux magnification and polarization angles.
Our algorithms can be easily realized in other programming languages such as FORTRAN,  C, and C++.  
The MATLAB version of KERTAP is parallelized using the MATLAB Parallel Computing Toolbox and the Distributed Computing Server.
The Python code was sped up using Cython and supports full implementation of MPI using `mpi4py' package. 
As an example, we investigate the inclination angle dependence of the observed polarization and the strong lensing magnification of  AGN X-ray emission.  
We conclude that it is possible to perform complex numerical-relativity-related computations using interpreted languages such as MATLAB and Python. 
\end{abstract}


\keywords{accretion, accretion disks --- polarization  --- gravitational lensing: strong --- quasars: supermassive black holes}

\maketitle

\section{Introduction}\label{sec:Intro}

Accretion disks are one of the most spectacular phenomena in modern astrophysics. 
The large luminosity of active galactic nuclei (AGN) and quasars (as large as $10^{47}$ erg/s) is believed to be the result of gas accreted by central supermassive black holes.
Up to $\sim$10\%  of the accreting mass can be emitted as radiation during the accretion process (much higher than 
nuclear fusion processes, about $\sim$0.5\%).
Besides emission in the infrared, optical, and UV bands, most AGN emit X-rays. 
Unlike optical emission, which can be explained by standard accretion disk theory (Shakura \& Sunyaev 1973), the physical mechanism of AGN  X-ray emission remains an enigma, despite decades of research effort (e.g., Krolik 2008).
According to the standard theory, the temperature of an AGN accretion disk is not hot enough to emit X-rays.
The X-ray emission is thought to be generated through inverse Compton scattering of disk photons with hot electrons in the so-called X-ray ``corona."
However, both the physics and geometry of this mysterious X-ray corona are not well understood.
Observations have revealed that photons of different frequencies correspond to emission regions of different sizes: the higher the frequency, the smaller the emission size and the closer it is to the central black hole.
Very recently, observations in quasar microlensing (Kochanek 2004; Blackburne et al.\ 2006, 2014, 2015; Morgan et al.\ 2008, 2012; Chartas et al.\ 2009; Dai et al.\ 2010; Chen et al.\ 2011, 2012; Mosquera et al.\ 2013; MacLeod et al.\ 2015) have constrained the AGN X-ray emission size to be of order 10 gravitational radii ($r_g\equiv GM/c^2$ where $M$ is mass of the central black hole). 
Within distances of a few $r_g,$ Einstein's general relativistic gravity theory plays a fundamental role in understanding accreting phenomena.
In particular, the X-ray emission is strongly lensed by the central black hole before it escapes the  gravity field and arrives at a distant observer. 
It is well known that photons follow null geodesics in curved spacetimes.
The spacetime around a rotating black hole is described by the Kerr metric (Kerr 1963).
The strong field environment of AGN X-ray emission invalidates the linear approximation as made in standard gravitational lensing theory (Schneider et al.\ 1992).
Consequently, numerical integration of the geodesic equations is necessary for many problems involving the Kerr metric.
Another well-known fact is that a photon's polarization vector is parallel transported along the photon's geodesic and its plane of polarization is rotated as the photon passes the rotating black hole, i.e., the so-called gravitational Faraday rotation (Balazs 1958; Plebanski 1960; Ishihara et al.\ 1988; Agol 1997;  Agol \& Krolik 2000; Frolov \& Shoom 2012; Yoo 2012).
Since X-rays are emitted in regions very close to the black hole, the observed X-ray polarization will be influenced by the strong gravity field.
Following the pioneering work of Stark, Connors, and Piran (Connors \& Stark 1977;  Stark \& Connors 1977;  Connors et al.\ 1980), X-ray polarization has now become a very promising probe of important parameters such as the black hole spin and the inclination angle of the accretion disk (Agol 1997; Agol \& Krolik 2000; Dov$\rm\check{c}$iak et al.\ 2004; Li et al.\ 2009; Schnittman \& Krolik 2009; Schnittman et al.\ 2013). 
It is therefore essential to include the effects of strong gravity when studying AGN X-ray emission.

Numerical ray-tracing in the Kerr metric is not new (Cunningham 1975; Luminet 1979; Rauch \& Blandford 1994; Bromley et al.\ 1997; Beckwith \& Done 2004; Fuerst \& Wu 2004; Schnittman \& Bertschinger 2004; Broderick 2006; Cadeau et al.\ 2007; Dolence et al.\ 2009; Psaltis \& Johannsen 2012; Chan et al.\ 2013; Chen et al.\ 2013a,\,b; Schnittman \& Krolik 2013).
However, public codes are relatively rare.
Exceptions are Dexter \& Agol (2009; a FORTRAN code, fast but mathematically convolved; see also Yang \& Wang 2013), Vincent et al.\ (2011; a C++ code, not parallelized), Kuchelmeister et al.\ (2012) and Chan et al.\ (2013; both are GPU-based ray-tracing code solving $2$nd order geodesic equations using CUDA).
Polarization and gravitational Faraday rotation have been studied extensively in the literature (Agol 1997; Agol \& Krolik 2000; Schnittman \& Krolik 2009); however, there seems to be no public code with a focus on X-ray polarizations.
Furthermore, there seems to be no ray-tracing code written in interpreted high-level languages such as MATLAB or Python.
We present  simple algorithms for computing the effect of strong lensing of AGN accretion disks by a Kerr black hole, including effects on polarization.
Our algorithms are realized in the high-level languages MATLAB and Python, and they are fully parallelized and can be easily realized in other languages such as FORTRAN and C++.

The plan of this paper is as follows.  
An introduction to the Kerr metric, the tetrad formalism, the strong lensing formalism, and the polarization formalism is given in Section~\ref{sec:basic}.
The graphic user interface (GUI) and numerical algorithms are discussed in detail in Section~\ref{sec:algorithm}.
In Section~\ref{sec:application}, we give an example of the use of our  MATLAB/Python program KERTAP (KErr Ray-Tracing And Polarization).
In Section~\ref{sec:discussion} we conclude.

\section{Theoretical Framework}\label{sec:basic}

Since the gas accreted by black holes carries angular momentum, the black holes powering accretion processes are expected to possess angular momentum themselves.
The proper metric describing the spacetime around a rotating black hole was discovered by Kerr in 1963 (Kerr 1963) and named after him.
We introduce the Kerr metric in this section, and discuss the other formalism needed in the strong lensing and polarization analysis of AGN X-ray emission.

\subsection{Kerr Metric}

In the Boyer-Lindquist coordinates, the Kerr metric can be written as
\bea\label{metric}
ds^2 
=-\alpha^2dt^2 +\frac{\Sigma^2}{\rho^2}\sin^2\theta(d\phi-\omega dt)^2+\frac{\rho^2}{\Delta}dr^2+\rho^2d\theta^2
\eea where $(t,r,\theta,\phi)$ are the four independent coordinate variables, and we have defined 
\bea\label{omega}
\rho^2&\equiv& r^2+a^2\cos^2\theta\cr
\Delta&\equiv&r^2-2Mr+a^2\cr
\Sigma^2&\equiv& \rho^2\left[(r^2+a^2)+\frac{2Mr}{\rho^2}a^2\sin^2\theta\right]\cr
\alpha^2&\equiv& \frac{\rho^2\Delta}{\Sigma^2}   \cr
\omega&\equiv& \frac{2aMr}{\Sigma^2}
\eea with constants $M$ and $a$ the mass and angular momentum (\ie J/M) of the Kerr black hole (we have taken $c=G=1$).
The geodesic motion of a photon (its 4-momentum satisfies $p^ap_a=0$, the index $a$ runs from 1 to 4, with 1 the time component) can be described by the group of 8 first order Hamilton equations (Arnold 1978)
\bea\label{H_equation}
\frac{dx^a}{d\xi}&=&\frac{\partial {\cal H}}{\partial p_a}  \cr
\frac{dp_a}{d\xi}&=&-\frac{\partial {\cal H}}{\partial x^a}
,\eea where $\xi$ is the affine parameter and ${\cal H}(x^a,p_b) \equiv \frac{1}{2}g^{ab}(x)p_ap_b$ is the Hamiltonian. 
The number of equations to integrate reduces to six since $p_t$ and $p_\phi$ are motion constants (the Hamiltonian $\cal H$ does not depend on $t$ or $\phi$). 
There is another motion constant discovered by Carter (1968),
\be
{\cal C}= p^2_\theta+\cos^2\theta\left(\frac{p^2_\phi}{\sin^2\theta}-a^2p_t^2\right).
\ee 
The number of ordinary differential equations (ODEs) used to solve the geodesic equations could have been reduced by one using this constant; however, we use $\cal C$ as an independent check of our ray-tracing code.
A Kerr black hole has two horizons (outer and inner) defined as $r_{\pm}= M\pm\sqrt{M^2-a^2}$.
The maximum angular momentum of a Kerr black hole was found in \cite{Thorne1974} to be $a_{\rm max}=0.998.$ 
We stop the ray-tracing when the geodesic is within $0.1\%$ of the outer horizon $r_{+}.$
For accretion disks (normally assumed in the equatorial plane, $\theta=\pi/2$), an important quantity is the so called innermost-stable-circular-orbit (ISCO) in the equatorial plane (Chandrasekhar 1983).
For example, $r_{\rm ISCO}=6\,r_g$ and $1.24\,r_g$ for a Schwarzschild black hole ($a=0$) and an extreme Kerr black hole.
We use $r_{\rm ISCO}$ as the inner cutoff of the accretion disk in this paper.

\subsection{Tetrad Formalism} 

In curved spacetime the frequency (or energy) of a photon depends on the spacetime point where the frequency is measured, and also on the 4-velocity $u^a$ of the observer measuring the frequency ($u^au_a=-1$).
If $p_a$ is the 4-momentum of the photon, the observed photon frequency $ \nu = -u^ap_a.$  
Let $\nu_{\rm o}$ and $\nu_{\rm e}$ be the photon's frequency measured by a distant observer and  an observer comoving with the light source (emitter).
We  define the redshift factor
\be\label{g_factor}
g\equiv\frac{\nu_{\rm o}}{\nu_e} =\frac{(-u^ap_a)_{\rm o}}{(-u^ap_a)_{\rm e}},  
\ee where  $u^a_{\rm o}$  and $u^a_{\rm e}$ are the 4-velocity of the distant observer  and the source. 

For the distant observer, we choose $u^a=\frac{1}{\alpha}(1,0,0,\omega),$ 
the so-called zero-angular-momentum observer (ZAMO; Bardeen et al.\ 1972) where $\omega$ and $\alpha$ are defined in Eq.~(\ref{omega}). 
A photon's momentum can be specified using the orthonormal tetrad of ZAMO instead of the coordinate basis, \ie 
\bea\label{zamo_coord_to_hat}
\left(\begin{array}{c}\hat{p}^t \\ \hat{p}^r \\ \hat{p}^\theta \\ \hat{p}^\phi\end{array}\right)=
\left(\begin{array}{cccc} \alpha & 0 & 0 & 0 \\0 & \frac{\rho}{\sqrt{\Delta}} & 0 & 0 \\0 & 0 & \rho & 0 \\ -\frac{\Sigma\sin\theta}{\rho}\omega & 0 & 0 &\frac{\Sigma\sin\theta}{\rho}\end{array}\right)
\left(\begin{array}{c}p^t \\ p^r \\ p^\theta \\ p^\phi \end{array}\right)
\eea
We will use Eq.~(\ref{zamo_coord_to_hat}) to compute the polarization measured by a distant observer (see Section~\ref{sec:polarization}).  
The four-velocity of the source is model-dependent.  
For accretion disks, we assume Keplerian flow for the accreting gas (Bardeen et al.\ 1972), \ie
\be\label{u_K}
u^a_K=\gamma(1, 0, 0,\Omega_K), 
\ee where 
\bea\label{Omega_K}
\Omega_K&\equiv&\frac{M^{1/2}}{r^{3/2}+a M^{1/2}}\cr
\gamma &\equiv& (-g_{11}-2 g_{14}\Omega_K -g_{44}\Omega_K^2)^{-1/2}.
\eea 
We similarly define an orthonormal tetrad in the comoving frame of the Keplerian flow,
and to change between tetrad and coordinate components we use
\be\label{hat_to_coord}
 \left(\begin{array}{c} p^t \\ p^r \\ p^\theta \\ p^\phi \end{array}\right)=
\left(\begin{array}{cccc} \gamma & 0 & 0 & \tau\lambda \\0 & \frac{\sqrt{\Delta}}{\rho} & 0 & 0 \\0 & 0 & -\frac{1}{\rho} & 0 \\ \gamma\Omega_K & 0 & 0 & \tau\end{array}\right)
\left(\begin{array}{c}\hat{p}^1 \\\hat{p}^2 \\ \hat{p}^3 \\ \hat{p}^4 \end{array}\right).
\ee
where
\bea
\lambda &\equiv& -\frac{g_{14}+g_{44}\Omega_K}{g_{11}+g_{14}\Omega_K}\cr
\tau &\equiv& (g_{11}\lambda^2+2 g_{14}\lambda +g_{44})^{-1/2}.
\eea
The above equations are useful since both the intensity and the polarization profile of the source emission are often given in the local rest frame of the source.  
In Section~\ref{sec:polarization}, we use Eqs.~(\ref{hat_to_coord})  to initialize the polarization propagator at the end of backward ray-tracing.

\subsection{Strong Lensing by Kerr Black Holes}
  
X-ray emission from an accretion disk  is strongly lensed by the central Kerr black hole as it leaves the source to arrive at a distant observer.
A lensed accretion disk appears very different from a unlensed one.
The frequency of the source emission is gravitationally and Doppler blue or redshifted depending on  the position of the emitter, the 4-velocity of the source, and  the distance and inclination angle of the observer. 
Consequently, the intensity profile of a lensed accretion disk can be very different from an unlensed one, and the observed flux can also be different from the unlensed case.

By definition, the monochromatic flux $F_{\nu_{\rm o}},$ measured by an observer O very far from the black hole is \citep{Mihalas78}
\be
F_{\nu_{\rm o}}\equiv\oint{I_{\nu_{\rm o}}\mu d\Omega_{\rm o}}= \int{I_{\nu_{\rm o}} d\Omega_{\rm o}},
\ee where $d\Omega_{\rm o}$ is the differential solid angle measured at the observer, and $\mu$ is the angle cosine between the light ray and the normal of the detector window.
We can safely drop the factor $\mu$ for sources at cosmological redshifts because these sources are observed with tiny solid angles. 
If we ignore Kerr strong lensing and Doppler shifts,  photons travel along straight lines, and there is no frequency redshift.
Consequently $I_\nu$ is a conserved quantity along the light path, 
\bea\label{F_flat}
F^{\rm unlensed}_{\nu_{\rm o}}\equiv\int{I^{\rm obs}_{\nu_{\rm o}}(\nhat)d\Omega_{\rm o}}=\frac{\cos\theta}{r_{\rm obs}^2}\int{I^{\rm source}_{\nu_{\rm o}}(\nhat)d{\cal A}}
\eea where $\nhat$ is the 3D photon direction at the observer, $d{\cal A}$ is the differential area element of the accretion disk, and we have changed the solid angle integration at the observer into a 2-D surface integral over the accretion disk.  
On the other hand, for Kerr lensing, we have 
\be\label{F_lensed}
F^{\rm lensed}_{\nu_{\rm o}}\equiv\int{g^3(\nhat) I^{\rm source}_{\nu_{\rm e}}(x^a(\nhat),p_a(\nhat)) d\Omega_{\rm o}},
\ee where $\nu_{\rm e}$ is the source frequency,  $g$ is the redshift factor of Eq.~(\ref{g_factor}),  $(x^a,p_a)$ is the photon's position and momentum at the emitter found by backward ray-tracing (see Section~\ref{sec:BRT}).  To obtain Eq.~(\ref{F_lensed})  we have simply used the fact that $I_{\nu}/\nu^3$ is conserved along each geodesic.
The strong lensing magnification of the specific flux is then
\be
\mu \equiv \frac{F^{\rm lensed}_{\nu_{\rm o}}}{F^{\rm unlensed}_{\nu_{\rm o}}}.
\ee 

\subsection{Backward Ray-tracing}\label{sec:BRT}

Since the emission region can be as close as a few gravitational radii from the black hole, the standard linearized gravitational lensing theory is not valid.
Our  strong lensing analysis in Kerr spacetime is based on backward ray-tracing.    
Given a source profile $I_{\nu}(x^a,p^b),$ the flux integral in Eq.~(\ref{F_flat}) can be easily computed either analytically or numerically.
However, it is not a trivial job to evaluate the lensed flux integral  Eq.~(\ref{F_lensed}) in which the integral is over the solid angle at the observer, but the integrand is evaluated at the source.
We do this through backward ray-tracing.
We choose a thin pencil beam (a sharp pentahedron) focused on the observer (the pigment core toward the black hole), and divide the solid angle space of this beam (large enough to contain all disk emission arriving at the observer) into a uniform grid of pixels.  
Through each pixel, we shoot one ray backward from the observer to the source. 
To compute the image area, we need only count the number of pixels whose light rays end at the accretion disk. 
To compute the observed flux, we weight pixels  by the integrands in  Eq.~(\ref{F_lensed}).  
In this way, our backward ray-tracing algorithm takes into account gravitational light deflection, Doppler or gravitational redshift, and area distortion simultaneously. 

\subsection{Polarization and Gravitational Faraday Rotation}\label{sec:polarization}

Because the polarization at the source is known and needs to be computed at the observer,  most current polarization propagators are based  on forward raytracing \citep{Schnittman09}.
Our polarization integrator is based on backward ray-tracing.
We assume that the polarization and limb darkening of the source emission are given as in the classical work of \cite{Chandra60} where the radiation transfer equation with polarization was solved assuming a scattering-dominated semi-infinite atmosphere.
Other initial polarization profiles, such as optically-thin, inverse Compton scattering-dominated atmosphere, can also be implemented.

At the end of the backward ray-tracing of each ray, we know the landing point on the accretion disk of a geodesic and its 4-momentum $p^a$.  
First, we compute the photon's 4 momentum components $\hat{p}^a$ in the comoving frame. 
We then compute the angle cosine, $\mu,$ between the photon wave vector and the disk normal measured by the comoving observer.
The angular dependent part of the intensity profile, $w(\mu)$ (see Eq.~(\ref{profile}) in Section~\ref{sec:application}), and the degree of polarization, $\delta_{\rm source}(\mu)$, can be computed by interpolating Table XXIV of \cite{Chandra60}.
We then compute the photon's polarization vector ${\bf E}$ in the local rest frame using the fact that ${\bf E}$ is parallel to the disk plane and normal to the photon's wave vector ($E\cdot p=0$), i.e., 
\be
\hat{E}^a = \frac{1}{\sqrt{(\hat{p}^2)^2+(\hat{p}^4)^2}}(0,-\hat{p}^4,0,\hat{p}^2).
\ee
Next we use Eq.~(\ref{hat_to_coord}) to compute the coordinate components $E^a $ of the polarization vector ${\bf E}.$
We then solve the parallel propagation equation of the polarization vector forward along the photon geodesic toward the distant observer using 
\be\label{E_propa}
\frac{d E^a}{d\xi} = -\Gamma^{a}_{\,bc}p^bE^c,
\ee where $\Gamma^{a}_{bc}$ are the Christoffel symbols.

At the end of the forward polarization propagating process, we are back to the observer with rotated polarization vector ${\bf E}_{\rm obs}$.
Here we first compute the tetrad components of ${\bf E}_{\rm obs}$ using Eq.~(\ref{zamo_coord_to_hat}).
We then work out the Stokes parameters corresponding to this ray,
\bea\label{Stokes}
2\chi &=&\arctan\left(-\frac{\hat{E}^\theta}{\hat{E}_{\phi}}\right)\cr
Q &=&\delta I_{\nu_o}\cos 2\chi\cr
U &=&\delta I_{\nu_o} \sin 2\chi
\eea where $\chi$ is the polarization angle, $\delta=\delta_{\rm source}$ is the degree of polarization which is conserved along a photon geodesic, and $I_{\nu_o}$ is the observed specific intensity at frequency $\nu_o,$ which is related to the source intensity by $I_{\nu_o}= g^3 I_{\nu_e}$.
To compute the  integrated degree of polarization and polarization angle, we add $Q,$  $U,$ and $I_{\nu_o}$ from each image pixel, obtaining $Q_{\rm total}$, $U_{\rm total},$ and $I_{\nu_o}^{\rm total}.$
Finally we use  Eq.~(\ref{Stokes}) again to compute $\chi_{\rm total}$ and $\delta_{\rm total}.$
Detailed algorithms are given in the next section.

\section{GUI And Numerical Algorithms of KERTAP}\label{sec:algorithm}

\subsection{Graphic User Interface}

\begin{figure*}
\includegraphics[width=0.8\textwidth,height=0.4\textheight]{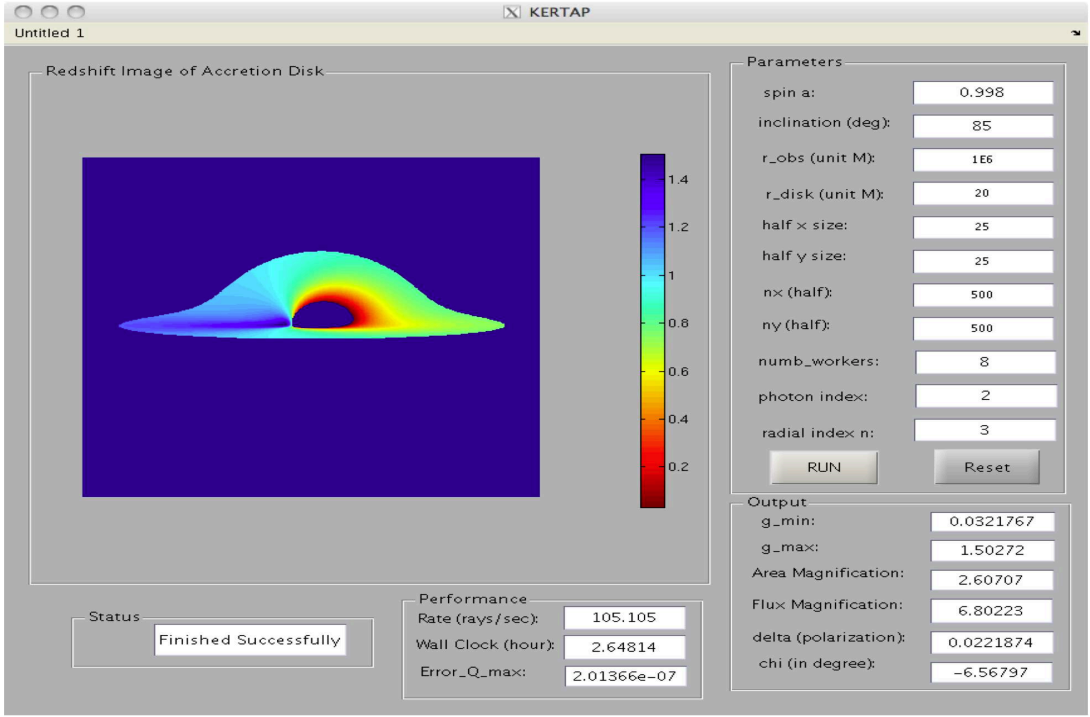}
\caption{GUI of KERTAP. 
Input parameters can be entered from the ``Parameters" panel. 
KERTAP will run after the user pushes the ``RUN" button. 
The status and performance of the program is shown in the ``Status" and ``Performance" panel, respectively.  
Important  outputs are given in the ``Output" panel.
The redshifted image of a lensed accretion disk will be plotted after a successful run. 
Pushing the ``Reset" button will clear the outputs and restore the default inputs.
A parallel computing job can be submitted to a remote cluster directly from the GUI on the user's desktop.
}
\label{fig:gui}
\end{figure*}

The MATLAB version of KERTAP contains a GUI (see Figure~\ref{fig:gui}).
It asks the user for the black hole parameters, the resolution desired, and the number of parallel workers for parallel computing.
The GUI contains a status panel indicating the status of the code: waiting for input, running, or finished.
If the code is run successfully, the GUI will generate the redshifted image of the accretion disk, and output important parameters such as the flux magnification and the degree and angle of the observed polarization. 
The performance of the code, such as the wall clock time used for ray-tracing, will also be recorded.

\subsection{Initialization Algorithm For Backward Ray-tracing}

Given an observer distance $r_{\rm obs},$ and the dimensions of the image rectangle (orthogonal to the line of sight at the source redshift), $x_{\rm size}$ and $y_{\rm size},$ the angular size of the cone focusing at the observer can be inferred.  
Given the resolutions in the x and y directions, the pixel size can be computed as $dx_{\rm grid}= x_{\rm size}/n_{\rm x}$ and $dy_{\rm grid}= y_{\rm size}/n_{\rm y}.$ 
For a pixel labeled by $(ix, iy)$ the photon's starting 8D phase-space coordinates are computed in {Algorithm 1}.

\begin{algorithm}[H]
\caption{	Initialization of one pixel (ix,iy) of the grid. 
		$\hat{p}^{r,\theta,\phi}$ are elements of the 3D photon momentum vector with respect to the ZAMO. 
		Function {\it p3\_to\_p4()} will normalize it assuming $\hat{p}^t=1,$ and then convert the ZAMO tetrad components to covariant coordinate components. 
		We choose  $t_{\rm obs}=0$ and $\phi_{\rm obs}=0$ without loss of generality. 
		}
\begin{algorithmic}
\State{$X_0=(0,r_{\rm obs},\theta_{\rm obs}, 0)$} \Comment{observer's 4-coordinates}
\State $\hat{p}^r= r_{\rm obs}$    \Comment{ must be positive}
\State $\hat{p}^\phi    = dx_{\rm grid}\times ix $
\State $\hat{p}^\theta = dy_{\rm grid}\times iy$
\State $p4= p3\_to\_p4 (\hat{p}^r, \hat{p}^\theta,\hat{p}^\phi)$  \Comment{p4 is covariant}
\State $Y_0 =(X_0, p4)$    \Comment{8-D starting point for RT }
\end{algorithmic}
\end{algorithm}

\subsection{Ray-Tracing Using MATLAB ODE45}\label{sec:ode45}

For each ray arriving at the observer, we solve the geodesic equations using MATLAB ODE solver {\it ode45}.
The syntax of {\it ode45} is 
\be\label{ode45}
[\xi,Y,\xi_e,Y_e,I_e] = ode45(@Y',\xi_{range},Y_0,options,...),
\ee where $Y'(\xi, Y)$ is an 8-D column vector containing the right hand side (RHS) of Eq.~(\ref{H_equation}),  $Y_0$ is the 8-D initial condition for {\it ode45}, and $\xi_{range}$ is the range of the independent variable (affine parameter) where we want to solve the ODEs. 
A subtle point of backward ray-tracing is that we do not  shoot rays toward the accretion disk (say, take $\hat{p}^r<0$), instead we must have $\hat{p}^r>0$ (toward the observer).
This point is important because the Kerr metric is not symmetric with respect to the reflection of the time coordinate (time reversal will change the direction of the black hole angular momentum).
The backward ray-tracing is realized by properly choosing $\xi_{range},$ \ie we define $\xi_{range}=[\xi_0, \xi_1]$ where $Y(\xi_0)=Y_0,$ and $Y(\xi_1)$ corresponds to the light emitting event ($\xi_1<\xi_0$, and $t_1< t_0$).    
In Eq.~(\ref{ode45}) {\it options} is a structure containing optional parameters that change the default integration properties, such as the relative and absolute error tolerance, and {\it events} to stop the integration.
For example, suppose we want to stop the ray-tracer when the geodesic is very close to the event horizon, say, $r<(1+\delta)r_{+}$ for some small $\delta$, we define a black hole hitting event as
\be
\rm flag_{BH}= (r\le (1+\delta)r_{+})-1.
\ee   
Since we want to know when the geodesic hits the equatorial plane, we define another event
\be
\rm flag_{Equator} = \cos \theta.
\ee
Then {\it ode45} will record the point where $\rm flag_{\rm BH}=0$ or $\rm flag_{\rm Equator}=0$ and stop the integration if the user requests to (as we did). 
These two events are defined in {\it disk\_events.m}. 
In the left hand side of Eq.~(\ref{ode45}), the index $I_e$ tells you which event happens, $(\xi_e, Y_e)$ stores the point where the event happens, and $(\xi, Y)$ stores the results of the integration up to the event point (\ie the photon geodesic).
Based on {\it ode45}, the ray-tracing part of KERTAP is easily realized in {Algorithm 2}.

\begin{algorithm}[H]
\caption{	Ray-tracing Algorithm. 
		MATLAB ode45 is  based on Runge-Kutta45. 
		}
\begin{algorithmic}
\For{ $ix = -nx:nx $ }
\For{ $iy = -ny:ny $} 
\State{Initialize pixel (ix,iy) giving $Y_0$ }\Comment{See {\bf Algorithm 1}}
\State{$[\xi,Y, ...]= ode45(@Y'(\xi), Y_0, options, ...)$}
\If{ ray goes to infinity}
\State{go to next ray}
\ElsIf{ hit on the horizon}
\State{$n_{\rm hole} = n_{\rm hole} +1$}
\State{go to next ray}
\ElsIf{hit on the equatorial plane}
\If{ hit on the accretion disk}
\State{$n_{\rm disk}=n_{\rm disk}+1$}
\State{compute redshift, write $(x^a,p_b)$, ...}
\State{go to next ray}
\EndIf
\EndIf 
\EndFor 
\EndFor
\end{algorithmic}
\end{algorithm}

\subsection{Polarization Propagator}

At the end of the backward ray-tracing of each ray, the polarization  propagator is initialized, and propagated forward to the observer, see {Algorithm 3}.

\begin{algorithm}[H]
\caption{	Polarization Propagating Algorithm. 
		In the following, {\it polarization()} computes the polarization degree $\delta_{source}$ and polarization vector $E^a_{source}$ on the accretion disk using $(x^a,p^a)_{source}$ obtained from backward ray-tracing.
		$\delta_{source}$ is interpolated from Table XXIV of \citet{Chandra60}.
		The polarization vector $E^a$ is propagated forward from the source to the observer using Eq.\,(\ref{E_propa}).  
              	 {\it polar\_obs()} computes the polarization angle $\chi$ at the observer.	
                  $\delta_{obs}=\delta_{source}$ since the degree of polarization is conserved along geodesics.
               	 }
\begin{algorithmic}
\For{ $ix = -nx:nx $ }
\For{ $iy = -ny:ny $} 
\State{$[\xi,Y,Y_e ]= ode45(@Y',  ...)$} \Comment{{\bf Algorithm 2}}
\If{ a good shot}
\State{$[\delta_{source},E_{source}] = polarization(Y_e,M,a)$}
\State{$E_{curr}= E_{source}$}
\State{$n_{point} = length(\xi$)}
\For{$j = 0, n_{point}-2$}\Comment{$E_{source}\rightarrow E_{obs}$}
\State{$d\xi = \xi_{j+1}-\xi_{j}$}
\For{$a = 1:4$}
\State{$dE^a =-\sum_{b,c=1}^{4}\left(\Gamma^a_{\>bc}p^bE^c_{curr}\right)d\xi$}
\State{$E_{next}^a= E_{curr}^a+dE^a$} 
\EndFor
\State{$E_{curr}= E_{next}$}
\EndFor
\State{$E_{obs}=E_{curr}$}
\State{$\chi = polar\_obs(E_{obs})$}
\EndIf 
\EndFor 
\EndFor
\end{algorithmic}
\end{algorithm}

\subsection{Post Processing}

The results of ray-tracing are stored in different data structures. 
For example, Image\_G and Image\_Y store the redshifts ($g=0$ for a missed shot) and 8-D phase space coordinates $(x^a,p_b)$ of the target points on the accretion disk for all successful shots.
Image\_P stores the polarization data $(w,\delta,\chi)$ where $w$ is the limb darkening factor from Table XXIV of \cite{Chandra60}.   
The strong lensing amplifications of the observed flux and image area  are computed in {Algorithm 4}.
The observed degree and angle of polarization  are computed using {Algorithm 5}.
Thanks to the data visualization tool intrinsic to MATLAB  and Python, the image of the lensed disk can be easily generated using functions such as {\it imagesc()} for MATLAB and {\it imshow()} for Python, and the gravitational Faraday rotated polarization vector field can be conveniently visualized using the function {\it quiver()} which is similarly defined for the two languages.
The redshifted image of an accretion disk will be plotted in the GUI after a successful run, and another two image windows (one for the same redshift image, the other for the intensity profile and polarization visualization) will pop up for the user to edit or save the figures in their desired formats (eps, jpg, etc).

\begin{algorithm}[H]
\caption{Strong lensing algorithm.  
               $dA$ and $d\Omega$ are the differential image area and solid angle for one pixel.
               The unlensed image area and flux are easy to compute, and are assumed to be known.}
\begin{algorithmic}

\State{$F^{lensed}=0$; $A^{lensed}=0$} \Comment{Initialization}
\For{ $ix = -nx:nx $ }
\For{ $iy = -ny:ny $} 
\If{ a good shot}
\State{$A_{lensed}=A_{lensed}+1$}
\State{$g=Image\_G(ix,iy)$}\Comment{read the redshift g}
\State{$Y=Image\_Y(ix,iy)$}\Comment{read 8-D $(x^a,p_b)$}
\State{$dF =g^3 I_{\nu_o/g}(Y)$}\Comment{$\nu_e=\nu_o/g$}
\State{$F^{lensed} = F^{lensed}+dF$}
\EndIf
\EndFor 
\EndFor
\State{$A^{lensed}=A^{lensed}\times dA$}
\State{$F^{lensed} = F^{lensed}\times d\Omega$}
\State{$\mu_{\rm area}= A^{lensed}/A^{unlensed}$}\Comment{Area amplification}
\State{$\mu_{\rm flux}= F^{lensed}/F^{unlensed}$}\Comment{Flux amplification}
\end{algorithmic}
\end{algorithm}

\begin{algorithm}[H]
\caption{Polarization synthesizing algorithm.  
               $\delta_{total}$ and $\chi_{total}$ are the integrated  degree and angle of polarization.}
\begin{algorithmic}

\State{$I_{total}=0;Q_{total}=0;U_{total}=0$} \Comment{Initialization}
\For{ $ix = -nx:nx $ }
\For{ $iy = -ny:ny $} 
\If{ IsImagePoint}
\State{$g=Image\_G(ix,iy)$}\Comment{read the redshift g}
\State{$r=Image\_Y(ix,iy,2)$}\Comment{read r coordinate}
\State{$[w,\delta,\chi]=Image\_P(ix,iy)$}
\State{$dI =wg^{\Gamma+2}/r^n$ }\Comment{See Eq.~(\ref{F_lensed2})}
\State{$dQ=\delta \times dI\times\cos(2\chi)$}
\State{$dU=\delta \times dI\times\sin(2\chi)$}
\State{$Q_{total}= Q_{total}+dQ$}
\State{$U_{total}= U_{total}+dU$}
\State{$I_{total}= I_{total}+dI$}
\EndIf
\EndFor 
\EndFor
\State{$\chi_{total}=0.5\tan^{-1}(U_{total}/Q_{total})$}
\State{$\delta_{total} = U_{total}/(I_{total}\sin(2\chi_{total})) $}
\end{algorithmic}
\end{algorithm}

\subsection{Python Version of KERTAP}

Despite the increased use of MATLAB in academia, for users who do not have access to a MATLAB cluster, we also provide a Python version of KERTAP.
The Python code uses exactly the same algorithms as the MATLAB code except that it does not have a GUI.
Furthermore, for the MATLAB code we used the ODE solver, {\it ode45} provided by MATLAB (already highly optimized and with stringent error control), whereas for the Python code we created a $5^{\rm th}$ order Runge--Kutta solver using the Cash--Karp methods.
The MATLAB {\it ode45} provides a very convenient `event' capturing mechanism to stop the ray-tracing at user defined events (such as when a ray crosses the event horizon, or punches through accretion disk, see Sec.~\ref{sec:ode45}),
whereas for the Python code, these events are coded into the ODE solver by hand.  
Consequently, to extend the code to include other events (e.g., higher order Kerr-lensing images) is slightly more difficult for the Python code than for MATLAB.

The MATLAB version of KERTAP was parallelized using the Parallel Computing Toolbox and the MATLAB Distributed Computing Server (MDCS).
A very useful feature of the MDCS server is that it allows parallel jobs to be submitted from the user's desktop directly to a distant parallel computing clusters, i.e., the user provides the desired parameters through the GUI of KERTAP,  and KERTAP automatically creates a job submission script and submits the job to the remote cluster dealing with popular job schedulers such as MOAB, SLURM, etc. 
The Python version of KERTAP was parallelized using the free {\it mpi4py} package which supports a full implementation of Message Passing Interface (MPI).
The pure Python version of KERTAP is slower than the MATLAB version by a factor of $\sim$2.
We speed up the Python code by adding C-type static type declarations to the pure Python code using {\it Cython} which results in a $\sim$100 times performance gain, see Fig.~\ref{fig:scalability}.
We provide our users with a pure Python version of KERTAP, a Cython version, and a parallel Cython version.
To run the Cython version of KERTAP, the Cython source file `mod.pyx' which contains most of the routines needs to be compiled into a Python extension module `cymod.so' before it can be imported as a Python package.
To run the Cython code on a cluster in parallel, a MPI compiler (e.g., gnu-openmpi) is needed.
Details of these implementations are included in the documentation to KERTAP.

\section{An Example Application}\label{sec:application}

We assume that the X-ray corona is a thin layer immediately above the AGN accretion disk, and moves with Keplerian flows (see Eqs.~(\ref{u_K}) and (\ref{Omega_K})).
We take $r_{\rm disk}=20\,r_g$ according to constraints obtained from quasar microlensing observations (Kochanek 2004; Blackburne et al.\ 2006, 2014, 2015; Morgan et al.\ 2008, 2012; Chartas et al.\ 2009; Dai et al.\ 2010; Mosquera et al.\ 2013; MacLeod et al.\ 2015), and use $r_{\rm ISCO}$ as the inner cutoff.
The image window for backward ray-tracing is of size $50\,r_g\times 50\,r_g.$
X-ray coronae of this type are motivated by the so-called sandwich model \citep{Haardt91,Haardt93}, and are often used in the literature \citep{RF2000}.
For a corona geometry such as a sphere, or a disk above the accretion disk, see Chen et al.\ (2013a).
Because quasar X-ray emission is observed to follow a power law, we assume the following form for the source intensity profile, 
\be\label{profile}
I_\nu(\nu,\mu, r) \propto \frac{1}{r^n}\frac{w(\mu)}{\nu^{\Gamma-1}}
\ee where $\mu$ is the cosine between the photon wave-vector and the upward disk normal measured by the comoving observer, and  $w(\mu)$ is the angular-dependence of  the intensity profile (we have assumed azimuthal symmetry for simplicity). 
To be more specific, we take $w(\mu)$ from \citet{Chandra60} (classical results from solving the radiation transfer equation with polarization assuming a scattering-dominated semi-infinite atmosphere).
The parameter $\Gamma$ is called the photon index, and $n$ specifies the radial steepness of the profile.   
We take $\Gamma=2.0$ (e.g., Chen et al.\ 2012), and $n=3$ (radially steep) or $0$ (radially flat).
We test for two black hole spins, a = 0.998 (Thorne 1974) or 0 (Schwarzschild black hole).
We take the observer distance as $r_{\rm obs} = 10^6 r_g$ and test for different inclination angles (from nearly face on to nearly edge on).
 
For the assumed simple geometry and intensity profile, the unlensed specific flux at frequency $\nu_o$ can be analytically integrated using Eq.~(\ref{F_flat})
\bea\label{F_flat2}
F^{\rm unlensed}_{\nu_o}&=&\frac{2\pi\mu_{\rm obs}w(\mu_{\rm obs})}{r_{\rm obs}^2}\frac{1}{\nu^{\Gamma-1}_o}\int_{r_{\rm ISCO}}^{r_{\rm disk}}{\frac{dr}{r^{n-1}}}\cr
&=& \frac{2\pi\mu_{\rm obs}w(\mu_{\rm obs})}{(n-2)r^2_{\rm obs}}\frac{1}{\nu^{\Gamma-1}_o}\left[\frac{1}{r^{n-2}_{\rm ISCO}}-\frac{1}{r^{n-2}_{\rm disk}}\right],
\eea where $\mu_{\rm obs}=\cos(\theta_{\rm obs}).$
We have dropped the unimportant constant in the definition of $I_{\nu}.$ 
For $n=2$, the $(n-2)$ factor on the denominator is absent, and the expression within the square brackets is replaced by $\log r_{\rm disk}/r_{\rm ISCO}$.
From Eq.~(\ref{F_lensed}) the lensed flux is 
\bea\label{F_lensed2}
F^{\rm lensed}_{\nu_o}&=&\frac{1}{\nu^{\Gamma-1}_o}\int{\frac{g^{\Gamma+2}w(\mu)}{r^n}d\Omega_o}\cr
&=&\frac{1}{r_{\rm obs}^2}\frac{1}{\nu^{\Gamma-1}_o}\left(\sum_{i,j}^{}{\frac{g^{\Gamma+2}_{i,j}w(\mu_{i,j})}{r^n_{i,j}}}\right)dA_{\rm grid},
\eea where the summation is over the accretion disk image, and $dA_{\rm grid}= dx_{\rm grid}\times dy_{\rm grid}$ is the area of one image pixel.
Note that $w(\mu)$ varies from pixel to pixel because of the general relativistic light bending, and the aberration effect. 
The magnification $\mu=F^{\rm lensed}_{\nu_o}/F^{\rm unlensed}_{\nu_o}$ can be computed using Eqs.~(\ref{F_flat2}) and (\ref{F_lensed2}) immediately (note that the term $1/(r_{\rm obs}^2\nu^{\Gamma-1}_o)$ cancels out).

\begin{figure}
\epsscale{1.25}
\plotone{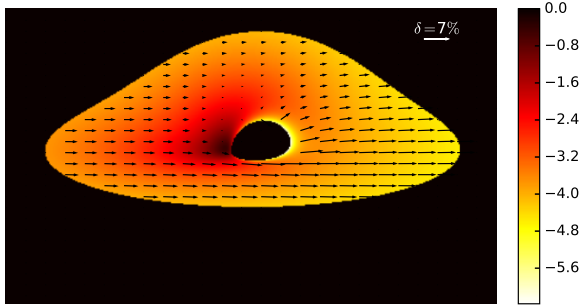}
\caption{Intensity and polarization profile of  a  X-ray disk of radius $r=20\,r_g$ strongly lensed by a Kerr black hole of spin $a=0.998$ observed at $\theta=75^\circ$. 
               The colorbar is given in log scale.
               Without the strong lensing effect, the image is an ellipse (with a hole), and the intensity profile is symmetric with respect to the y-axis. 
               The lensed disk is warped upward, and the intensity is concentrated in a small region on the left hand side of the disk (approaching the observer).
               The polarization should be horizontally aligned and constant in value ignoring spacetime curvature ($\delta =4.6\%$ for $\theta= 75^\circ$, see Chandrasekhar 1960).
               The strong lensing of the Kerr black hole reduces the degree of polarization significantly ($\sim$1.6\%).
               The orientation of the polarization can also be rotated near the black hole.
                 }
\label{fig:polarization}
\end{figure}

\begin{figure}
\epsscale{1.25}
\plotone{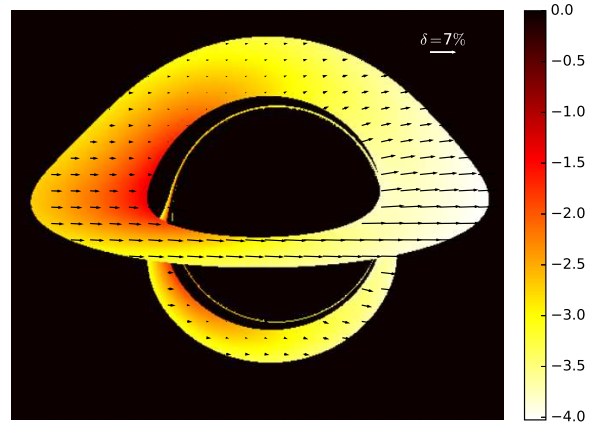}
\caption{    Intensity and polarization profile of  a  X-ray disk of radius $r=10\,r_g$ strongly lensed by a Kerr black hole of spin $a=0.5$ observed at $\theta=75^\circ$. We include up to $4^{\rm th}$ order Kerr-lensing images. 
            }
\label{fig:polar_order4}
\end{figure}

\begin{figure*}
\begin{center}$
\begin{array}{cc}
\includegraphics[width=0.45\textwidth,height=0.5\textheight]{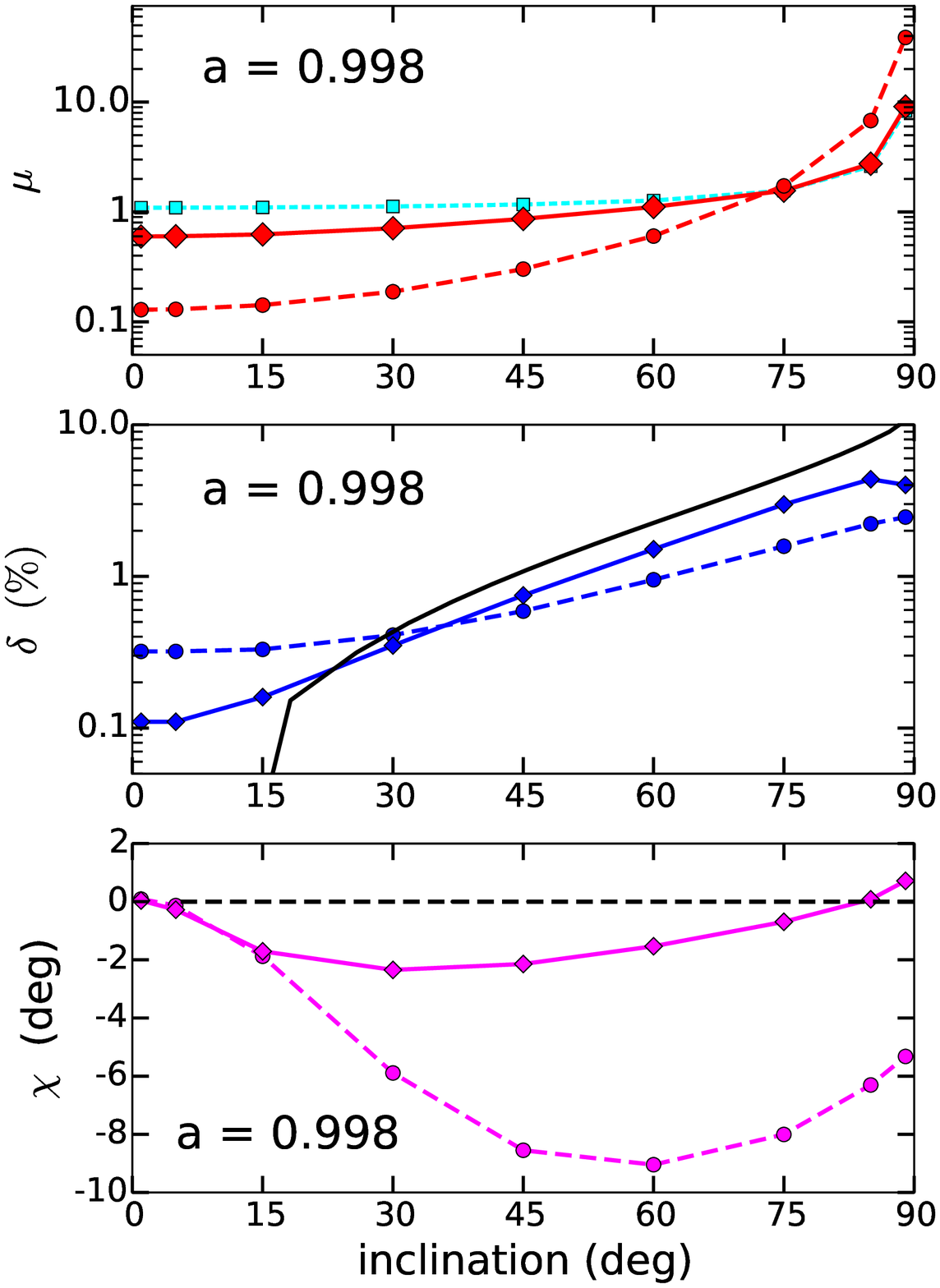}
\hspace{10pt}
\includegraphics[width=0.45\textwidth,height=0.5\textheight]{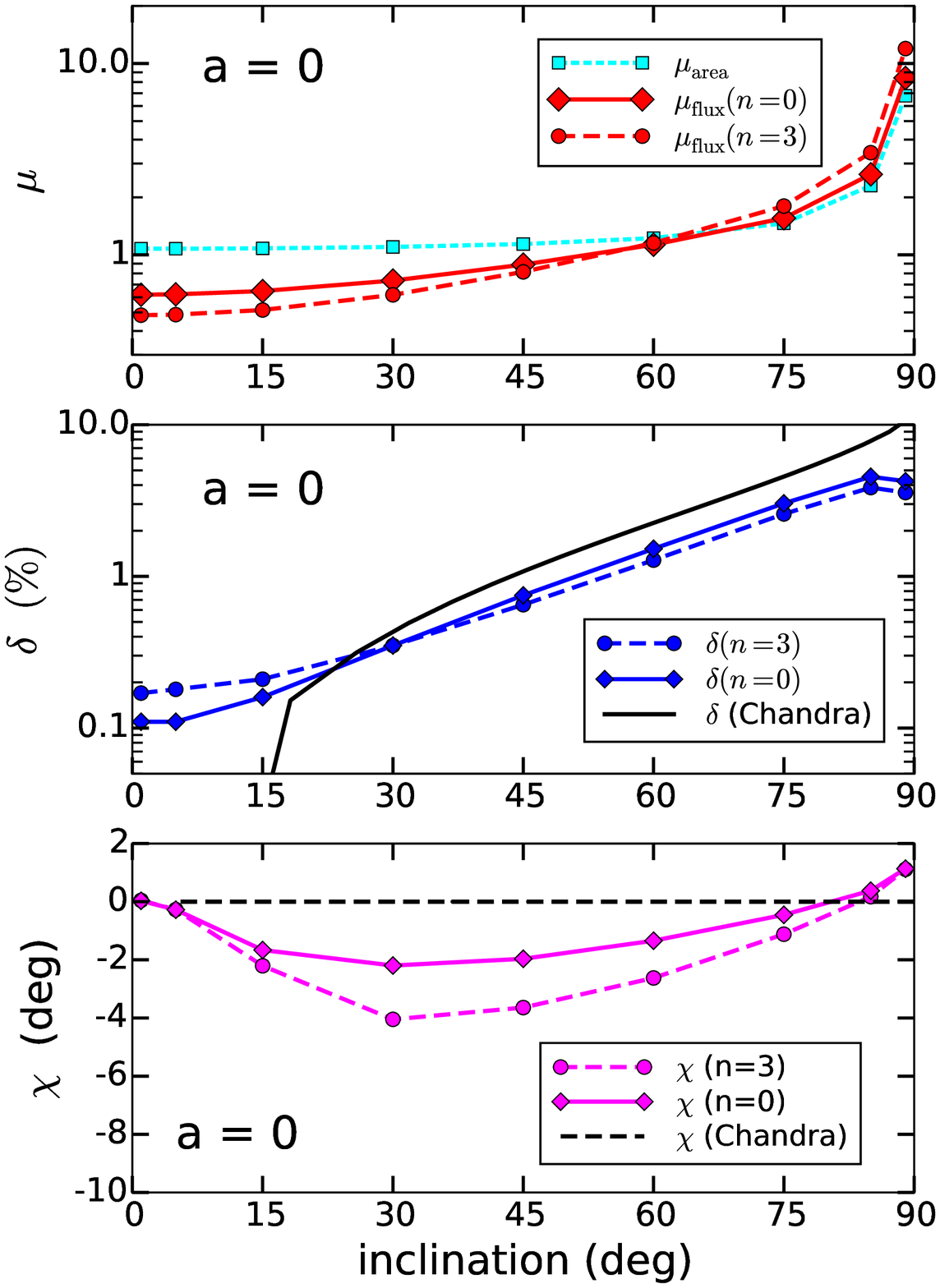}
\end{array}$
\end{center}
\caption{Inclination dependence of Kerr strong lensing and polarization.
We test for two black hole spins $a=0.998$ and $0$ (the first and second column), and two radial profiles $n=3$ (steep) and $n=0$ (flat).
The first row plots the strong lensing magnification of the image area and total flux.
The second and third rows plot observed degree and angle of polarization, respectively. 
The black lines in the second and the third row are classical \citet{Chandra60} results.}
\label{fig:lensing}
\end{figure*}

In Figure~\ref{fig:polarization} we plot the lensed intensity profile and distribution of polarization for the X-ray disk observed at inclination angle $75^\circ.$
Figure~\ref{fig:polar_order4} is similar to Fig.~\ref{fig:polarization}, but is for black spin $a=0.5$  and we have included up to $4^{\rm th}$ order Kerr-lensing images.
Kerr strong lensing changes both the shape and the intensity profile of the disk significantly. 
The intensity is strongly concentrated in a small region (in dark red) near the black hole and to the left of the accretion disk (where the rotating source is approaching the observer).
As for polarization, the degree of polarization to the left of the black hole is on average smaller than the classical value ($\sim$4.6\% for $\theta=75^\circ$; Chandrasekhar 1960) because the angle between the photon wave vector and the disk normal measured by the comoving observer is smaller than the inclination angle.
Therefore, $\delta(\mu)$ is smaller, see the black lines in the second row of Figure~\ref{fig:lensing} for the inclination dependence of $\delta$.
The contrary holds for the right hand side of the accretion disk. 
On the other hand, the direction of polarization can change significantly because of the gravitational Faraday rotation.
This effect is more significant for source regions closer to the black hole where the observed polarization angle $\chi$ can be positive or negative (depending on the actual source location) instead of horizontally aligned as in \citet{Chandra60}.

The strong lensing magnification of the image area and specific flux, and the integrated (or averaged) polarization degree and angle are plotted in Figure~\ref{fig:lensing}. 
We did not consider higher order images here since the accretion disk is believed by many to be optically thick.
We plot the inclination dependence of $\mu_{\rm area},$  $\mu_{\rm flux},$ $\delta,$ and $\chi$ for spin $a=0.998$ and $0,$ and for steep and flat radial profile ($n= 3$ or 0).
The magnification increases with inclination angles and is more significant for steeper profiles and for larger spins (smaller $r_{\rm ISCO}$) both of which amount to more concentrated emission closer to the black hole.
The fact that for small inclination angles (nearly face on), the observed degree of polarization is greater than the classical result is mainly caused by gravitational light bending ($\delta=0$ for face on case, see Chandrasekhar 1960).
For moderate to high inclination angles, the emission is strongly focused in a small region with a low degree of polarization, see Figure~\ref{fig:polarization}. 
Consequently, special and general relativity effects tend to reduce the observed degree of polarization confirming earlier results (Connors et al.\ 1980).

Another interesting point is that for a fixed emission size (e.g., $r_{\rm disk}= 20\,r_g$ as used in this paper) with the intensity profile Eq.~(\ref{profile}), the effects on polarization, \ie the polarization profile, the integrated degree and the angle of polarization, are achromatic.
This is because both the inputs of the polarization propagator, \ie $\delta(\mu)$ and $w(\mu)$ from \cite{Chandra60}, and gravitational Faraday rotation are wavelength independent,  and the 1/$\nu_o^{\Gamma-1}$ factor cancels out when averaging over the image of the disk  when computing $\chi_{\rm total}$ and $\delta_{\rm total}$, see Eqs.~(\ref{Stokes}), (\ref{F_flat2}), and (\ref{F_lensed2}).
This apparent discrepancy with \citet{Schnittman09} is caused by the fact that \citet{Schnittman09} have assumed thermal emission for X-rays emitted by the accretion disk around a stellar mass black hole, and therefore, the X-ray emission of higher frequencies are emitted by regions closer to the black hole (the effective temperature of the disk is higher closer to the black hole), whereas we have assumed the same power-law $\propto \nu^{-(\Gamma-1)}$ at each radius for X-ray emission by AGN.
If we assume that hard X-rays are emitted by a region smaller than that of the soft X-rays and closer to the black hole as suggested by \citet{Chen11}, say, take $r_{\rm disk}= 10\,r_g,$ then, even under the same power-law assumption, the integrated degree and angle of polarization should  differ from current results, \ie the results will become chromatic.
  
\begin{figure}
\epsscale{1.1}
\plotone{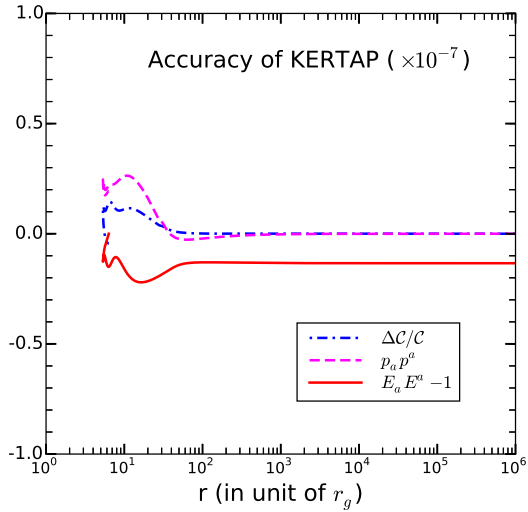}
\caption{ Accuracy of KERTAP. 
               A typical ray is shot backward toward the black hole, hits on the accretion disk, after which the polarization vector is forward propagated to the observer.
               The dot-dashed blue curve shows the relative error in Carter's constant.
               The dashed magenta curve shows the norm of the photon 4-momentum vector.
               The solid red line shows error in the norm of the forward parallel-propagated (space-like) photon polarization vector.
                 }
\label{fig:accuracy}
\end{figure}

\begin{figure}
\epsscale{1.1}
\plotone{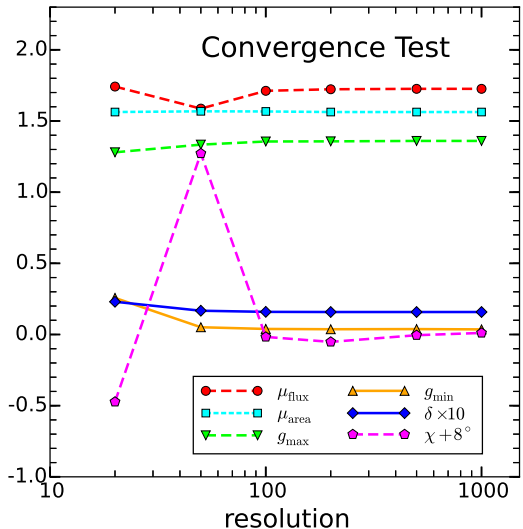}
\caption{Convergence test of KERTAP. 
We plot the resolution dependence of the maximum and minimum redshift, $g_{\rm max}$ and $g_{\rm min}$, the strong lensing magnification of the image area and specific flux, $\mu_{\rm area}$ and $\mu_{\rm flux}$, and the observed degree and angle of polarization.
The data are tabulated in Table~\ref{tab:converge}. 
For an X-ray emitting disk of size $20 r_g,$ a $500\times 500$ grid for an image rectangle of size $50r_g\times 50r_g$ is accurate enough for the Strong lensing calculations presented in this paper.
A grid of this size takes $\sim$45 minutes to run on a single computer with 8 MATLAB workers.}
\label{fig:converge}
\end{figure}

\begin{figure}
\epsscale{1.1}
\plotone{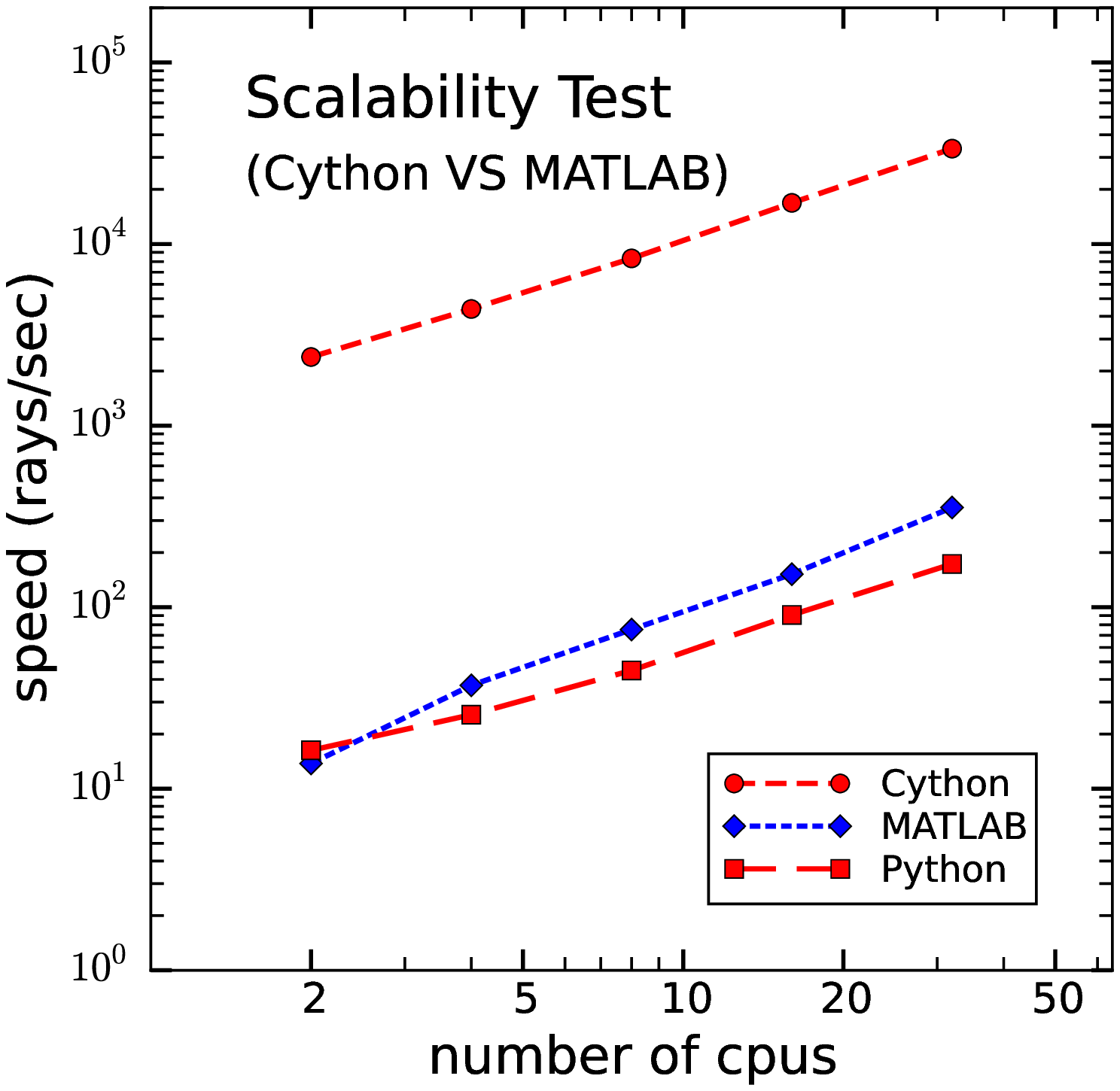}
\caption{Scalability of the parallel Python and MATLAB code.
The test was run for the example in Fig.~\ref{fig:polarization} with $x_{\rm size}=25\,r_g,$ $y_{\rm size}=15\,r_g,$ and resolution $800\times480.$
The rays were shot from $r=10^6\,r_g$ from the black hole, with a stringent error control (relative error $\epsilon=10^{-11}$ for the RK45 ode solver). 
The Python code was parallelized using the `mpi4py' package whereas the MATLAB code was parallelized using the MDCS. 
Both codes show good scalability.
The Python code was sped up by two orders of magnitude using Cython.
The curve is steeper for the MATLAB code when the number of CPUs is small, this is simply caused by the fact that among the N-workers allocated for the parallel job, one is selected as the master worker, and only N-1 workers are doing the ray-tracing. 
The number of parallel MATLAB workers is restricted by the number of MDCS licenses available on a MATLAB cluster, whereas the number of Python workers is only restricted by the size of the cluster.  
}
\label{fig:scalability}
\end{figure}

The accuracy of KERTAP is very good.
We have used Carter's constant as an independent check of accuracy.
At the end of the long ray-tracing ($r_{\rm obs}=10^6 r_g$), the relative error of Carter's constant is never worse than  $\sim$$10^{-7}$ for all cases tried by the authors.
We also checked the accuracy of KERTAP using the norm of the photon's 4-momentum vector and the polarization vector $E^a$, see Fig.~\ref{fig:accuracy} for results of a typical ray.
The results of the MATLAB code are highly consistent with the Python code, e.g., the norm of the difference between the redshift images generated using the Python and MATLAB code is of order $\sim$$10^{-6}$ and there were no rays which hit on the accretion disk or enter the horizon in one case that did not hit on the disk or entering the horizon in the other.
A convergence test is shown in Table~\ref{tab:converge} and Figure~\ref{fig:converge}.
The convergence of the KERTAP is pretty good, e.g.,  the amplification to image area and specific flux, and the polarization degree and angle all converge at a grid of moderate size $500\times 500$. 
In Fig.~\ref{fig:scalability} we test the scalability of KERTAP. 
Both the MATLAB (parallelized using MCDS) and Python (parallelized using {\it mpi4py}) code scale roughly linearly with the number of CPUs.
Consequently, the performance of KERTAP can be significantly improved by parallel computing.
However, the performance improvement of the MATLAB code through parallel computing was restricted by the number of MDCS licenses available in a user's MATLAB cluster,  whereas the Python code is only restricted by the size of the cluster (number of CPUs available for the MPI job).
It is important to note that since MATLAB2014b, the Parallel Computing Toolbox allows up to 512 parallel workers (this number is not constrained by the number of MDCS licenses available).
Consequently, significant speedup can still be achieved by users without a MATLAB cluster but with a single node with many cores.

\section{Discussion}\label{sec:discussion}

We have developed KERTAP, the first ray-tracing code using interpreted languages MATLAB and Python, for studying strong gravitational lensing in Kerr spacetime including polarization.\footnote{The code is available at https://bitbucket.org/binchen14/kertap }
The key ingredients of KERTAP are:  a GUI (for the MATLAB version),  backward ray-tracing realized by $5^{\rm th}$ order Runge-Kutta method, a polarization propagator and synthesizer, and the formalism and algorithms for computing strong lensing effects of a Kerr black hole centered on an accretion disk.
The backward ray-tracing part of KERTAP most resembles that of Schnittman \& Bertschinger (2004) and Vincent et al.\ (2011).
\citet{Vincent2011} focuses on radiation transfer while this paper focuses on observable effects of Kerr strong lensing and polarization. 
The independence of the geodesic equations from the polarization propagation equations makes it possible to solve the geodesic equations (backward) first, then propagate  the polarization vector forward. 
The idea of backward ray-tracing followed by forward polarization propagating is new, to the best of our knowledge.
Polarization and gravitational Faraday rotation have been studied extensively in the literature (e.g., Agol 1997; Agol \& Krolik 2000; Schnittman \& Krolik 2009).
However, public code explicitly dealing with polarization is not yet available. 
Published papers investigating the X-ray polarization of accretion disks are basically all doing forward ray-tracing (with or without Monte-Carlo radiation transfer; see, e.g., \citet{Schnittman09}).
Since X-rays are emitted by sources very close to the central black hole and moving with highly relativistic flows,  in order to compute the observable property measured by a distant observer at a given inclination angle, at each point on the accretion disk and for each direction, many forward rays have to be traced (because no one knows which ray will go where).
This increases the forward ray-tracing time significantly, and the results are usually noisier than those from backward ray-tracing (e.g., compare Figure~1 of \cite{Schnittman09} with Figure~\ref{fig:polarization} of this paper).  
    
The idea of performing very computationally intensive jobs such as ray-tracing in Kerr spacetime using interpreted languages such as MATLAB or Python might sound odd at the first. 
It is hard to expect a dynamic language such as Python to compete in performance with FORTRAN or C++.\footnote{This does not imply that the back and forth ray-tracing algorithms presented in this paper are intrinsically slow.} 
In fact, the single core (MATLAB or pure Python) version of KERTAP performs poorly compared to code written in FORTRAN and C++, and to GPU-accelerated CUDA code (see, e.g., Dexter \& Agol 2009; Chan et al.\ 2013).
However, this does not mean that interpreted languages such as Python or MATLAB will/should not be used in CPU intensive astronomical computations such as Kerr ray-tracing. 
The performance of these languages has been improved significantly over the past decade, in particular, through the development of application programming interfaces (APIs) with low-level languages such as C/FORTRAN (e.g., the CPython API),  multi-CPU parallel computing (mpi4py for Python; MDCS for MATLAB), and by supporting GPUs.
For example, both Python and MATLAB support GPU programming, e.g., PyCUDA\footnote{http://mathema.tician.de/software/pycuda/} for Python.   
The performance of the pure Python version of KERTAP was improved by two orders of magnitude using Cython.
The disadvantage in speed is compensated by parallel computing. 
In MATLAB, this is achieved by using the Parallel Computing Toolbox on local machines with multiple cores, or by using the Distributed Computing Server over a cluster.
For Python, the Cython version of KERTAP supports full implementation of MPI.
For example, we have shown in this paper that  it is feasible to study strong lensing of X-ray emission of accretion disks around Kerr black hole using MATLAB/Python. 
The strong lensing analysis of a X-ray emitting  disk of radius $20\,r_g$ can be done with high accuracy in $\sim$40 minutes by running  the Matlab version of KERTAP on a desktop computer with multiple cores ($\lesssim 1$ minute using the Cython code).
The computing time can be further reduced on a high performance computing cluster given the massively parallel nature of ray-tracing and the good scalability of KERTAP (see Fig.~\ref{fig:scalability}).
The MATLAB GUI of KERTAP allows a ray-tracing job to be created on a local machine (e.g., the user's laptop), submitted to a remote high-performance computing cluster with the results sent back to the user upon a successful run, the most desired form of interactive high performance computing.\footnote{Refer to http://rcc.fsu.edu/software/MATLAB for the details of submitting parallel MATLAB jobs remotely.}    
If the reader desires, the strong lensing and polarization algorithms can be easily realized in FORTRAN or C++.

The advantage of coding using high-level languages such as Python and MATLAB is multifold.  
First, the object-oriented nature of MATLAB makes it very easy to build graphic user interfaces.
For example, the GUI as shown in Figure~\ref{fig:gui} is written using the MATLAB GUI building tool ``guide."
Secondly, it is much easier to write code using MATLAB or Python than FORTRAN or C/C++.
Programming in MATLAB/Python greatly shortens the whole cycle of a scientific project, i.e., from prototyping and debugging the code, to production runs, to visualization of the scientific results and generating high quality plots.
For example, this whole paper can be done using a single language (either MATLAB or Python). 
Our code is very short.
The kernel of the MATLAB version of KERTAP is only about 600 lines.
The kernel of the Python version of KERTAP is about 200 lines longer (including the Runge-Kutta solver).
 The reader should experience no difficulty in understanding our code after reading this paper.
Thirdly, since MATLAB was designed to be a data analysis and visualization tool (and thanks to free Python packages such as `matplotlib'), the post-processing of ray-tracing data using MATLAB or Python is almost trivial.
For example, the piece of code generating the redshifted image of the accretion disk (see Figure~\ref{fig:gui}) is only a few lines. 
Users do not have to write separate codes visualizing their results using software such as IDL or Mathematica. 
Our code would be very useful for scientists new to the Kerr geometry or in need of polarization calculations, or to those people who need a second code to compare with their own.
KERTAP is the first public code explicitly dealing with strong lensing of X-ray polarization.
For less complicated computing jobs, our MATLAB/Python code can be used directly or after some modification. 
For complicated computing jobs, KERTAP might still be useful if the user's institute has a parallel computing cluster since KERTAP is fully parallelized.  
We conclude that it is possible to perform complex numerical-relativity-related computations using interpreted languages such as MATLAB and Python.

\section{Acknowledgements}

NSF AST-0707704 and support for Program number  HST-GO-12948.04-A was provided by NASA through a grant from the Space Telescope Science Institute, which is operated by the Association of Universities for Research in Astronomy, Incorporated, under NASA contract NAS5-26555.
X.D.\ acknowledges the financial support from the NASA ADAP programs NNX11AD09G, NNX15AF04G and NSF grant AST-1413056.
We thank the anonymous referee for a careful review of this work.
We thank Jeremy D. Schnittman for comments.
B.C. thanks Peter Beerli and Paul van der Mark for discussions on parallel computing using Python and Shiyun Tang for help in generating one plot.
The performance analysis was conducted on the HPC cluster at the Research Computing Center at the Florida State University.

\LongTables
\begin{deluxetable}{ccccccc}
\tablecolumns{7}
\tablewidth{10pt}
\tablecaption{ 
Convergence and performance of KERTAP. 
We run KERTAP on a single computer with eight MATLAB parallel workers. 
Accretion disk size $r_{\rm disk}=20\,r_g.$
The size of the image window is $50\,r_g\times 50\,r_g.$ 
Black hole spin $a=0.998.$ 
Photon index  $\Gamma=2.0$ and radial profile $n=3.$
Inclination angle $\theta=75^\circ.$ 
Observer distance $r_{\rm obs}=10^6\,r_g.$
\label{tab:converge}}
\tablehead{
\colhead{Resolution} &
\colhead{$\mu_{\rm area}/\mu_{\rm flux}$} &
\colhead{ $g_{\rm min}/g_{\rm max}$ }&
\colhead{ $\delta$ } &
\colhead{ $\chi$  }&
\colhead{ Wall time } &
\colhead{Speed} \\
\colhead{} &
\colhead{} &
\colhead{} &
\colhead{} &
\colhead{(deg)} &
\colhead{(hr)} &
\colhead{(ray $\rm s^{-1}$)}
}
\startdata
$20  \times 20      $   &   1.563/1.742    & 0.257/1.280  & 0.0230 & -8.473   &  0.004   &   27.33   \\
$50  \times 50      $   &   1.568/1.587    & 0.051/1.334  & 0.0167 & -6.728   &  0.011   &   66.15   \\
$100\times 100    $   &   1.567/1.712    & 0.039/1.356  & 0.0159 & -8.017   &  0.034   &   83.65   \\
$200\times 200    $   &   1.563/1.723    & 0.037/1.357  & 0.0158 & -8.052   &  0.120   &   93.07    \\
$500\times 500    $   &   1.563/1.726    & 0.038/1.360  & 0.0158 & -8.005   &  0.713   &   97.75    \\
$1000\times1000 $   &   1.563/1.726    & 0.036/1.360  & 0.0158 & -7.989   &  2.881   &   96.62    	\\
\enddata
\end{deluxetable}

\end{document}